\pdfoutput=1
\documentclass[journal=jpccck,manuscript=article]{achemso}
\setkeys{acs}{usetitle = true}

\usepackage[version=3]{mhchem} 

\newcommand*{\citen}[1]{
  \begingroup
    \romannumeral-`\x 
    \setcitestyle{numbers}%
    \cite{#1}%
  \endgroup   
}

\author{Zhiming Wang}
\author{Xianfeng Hao}
\author{Stefan Gerhold}
\author{Zbynek Novotny}
\affiliation 
{Institute of Applied Physics, Vienna University of Technology, Vienna, Austria}

\author{Cesare Franchini}
\affiliation  
{Faculty of Physics $\&$ Center for Computational Material Science, University of Vienna, Vienna, Austria}

\author{McDermott Eamon John Gordon}
\affiliation  
{ Institute of Materials Chemistry, Vienna University of Technology, Vienna, Austria}

\author{Karina Schulte}
\affiliation  
{MAX IV Laboratory, Lund University, Lund, Sweden}

\author{Michael Schmid}
\author{Ulrike Diebold}
\affiliation [Vienna University of Technology]
{Institute of Applied Physics, Vienna University of Technology, Vienna, Austria}
\email{diebold@iap.tuwien.ac.at}

\title[\texttt{achemso} demonstration]
{Water Adsorption at the Tetrahedral Titania Surface Layer of SrTiO$_3$(110)-(4~$\times$~1) }

\begin{document}
\begin{abstract} 
The interaction of water with oxide surfaces is of great interest for both fundamental science and applications. We present a combined theoretical [density functional theory (DFT)] and experimental [Scanning Tunneling Microscopy (STM), photoemission spectroscopy (PES)] study of water interaction with the two-dimensional titania overlayer that terminates the SrTiO$_3$ (110)-(4~$\times$~1) surface and consists of TiO$_4$ tetrahedra. STM, core-level and valence band PES show that H$_2$O neither adsorbs nor dissociates on the stoichiometric surface at room temperature, while it dissociates at oxygen vacancies. This is in agreement with DFT calculations, which show that the energy barriers for water dissociation on the stoichiometric and reduced surfaces are 1.7 and 0.9 eV, respectively. We propose that water weakly adsorbs on two-dimensional, tetrahedrally coordinated overlayers.
\end{abstract}

\section{1. Introduction}

The discovery of photochemical water splitting on SrTiO$_3$ with no external bias under UV irradiation has motivated much research into the interaction of water with this material.\cite{Mavroides:apl76, Wrighton:jacs76} More recent reports of overall water splitting on SrTiO$_3$ with a NiO co-catalyst has renewed this interest.\cite{Townsend:12acsnano, Townsend:12ees} A fundamental question is simply whether water adsorption is molecular or dissociative.\cite{Henrich:77ssc, Ferrer:1980ss, Webb:1981ss, Egdell:1982cpl, Cox1983247, Eriksen:1987sap, Brookes:1987ssc, Brookes:1987ps, Wang:2002jvsta} For SrTiO$_3$(001), photoemission spectroscopy (PES), high-resolution electron energy loss spectroscopy (HREELS) and temperature programmed desorption (TPD) studies show that water does not adsorb on the stoichiometric surface at room temperature (RT) although molecular water adsorption has been observed below 150 K. However, dissociative adsorption was observed for water on both, Ar$^+$ bombarded and vacuum-fractured SrTiO$_3$(100) surfaces.\cite{Eriksen:1987sap, Brookes:1987ssc, Brookes:1987ps} Theoretical calculations are in agreement with experimental results, predicting molecular water adsorption on the stoichiometric SrTiO$_3$(100) surface.\cite{Azad:2005jpcb, Baniecki:2009jap, Guhl:prb10}

In this context it is important to note that SrTiO$_3$(100) forms a wide variety of reconstructions, which depend strongly on the preparation conditions and sample history.  Various groups report different results,\cite{Bonnell:2008rpp} thus it is not always straightforward to connect water adsorption experiments to the actual surface structure.  Recently the SrTiO$_3$(110) surface has received significant attention.\cite{Russell:2008prb, Enterkin:natm10, Wang:prb11, Li:prl11, Biswas:apl11} It was found that SrTiO$_3$(110) surface can be prepared reproducibly and reversibly with a variety of surface structures.\cite{Russell:2008prb, Wang:prb11} The ($n$~$\times$~1) ($n$ = 3 - 6) series of reconstructions was solved by transmission electron diffraction and direct methods, and confirmed and refined by density functional theory (DFT) calculations and scanning tunneling microscopy (STM).\cite{Enterkin:natm10, Li:prl11} Thus a reliable structural model is available for this surface.  

The SrTiO$_3$(110) surface is polar, as an SrTiO$_3$ crystal can be considered as a stack of equidistant (SrTiO)$^{4+}$ and (O$_2$)$^{4-}$ planes along the [110] direction.\cite{Bottin:ss05} Generally, polar surfaces are considered more reactive than non-polar ones.\cite{Noguera:jpcm08, Shin:nl09} In this case, however, the polarity is compensated via the formation of a (4~$\times$~1) reconstruction with a nominal stoichiometry of (Ti$_{1.5}$O$_4$)$^{2-}$.  The reconstruction consists of six- and ten-membered rings of corner-shared TiO$_4$ tetrahedra residing directly on the bulk-like SrTiO$_3$, which consists of octahedrally-coordinated Ti (see Fig. 1a). The surface reconstruction can be tuned by varying the surface stoichiometry,\cite{Wang:prb11, Wang:apl13} forming a homologous series of ($n$~$\times$~1) ($n$ = 3 - 6) with a variable number of tetrahedra per ring.\cite{Enterkin:natm10, Li:prl11} Recently, we reported that quasi-long-range ordered antiphase domains are formed on the (4~$\times$~1) surface.\cite{Wang:prl13} The domain boundaries are decorated by defect pairs consisting of a Ti$_2$O$_3$ vacancy cluster and a Sr adatom; the presence of these pairs preserves the polarity compensation. 

In recent reports, periodically-arranged, tetrahedrally coordinated MeO$_4$ (Me = Ti, Si) units have emerged as a common feature on several oxide surfaces.\cite{Enterkin:natm10, Li:prl11, Lazzeri:prl01, Blanco-Rey:prl06, Marks:ss09, Warschkow:prl08} For example, such units form one-dimensional rows at the anatase TiO$_2$(001)-(1~$\times$~4) and rutile TiO$_2$(110)-(1~$\times$~2)-Ti$_2$O$_3$ surfaces.\cite{Lazzeri:prl01, Blanco-Rey:prl06} For anatase (001) a high reactivity towards water adsorption was reported;\cite{Gong:jpcb06, Blomquist:jpcc08} this surface was also identified as the most active one in photocatalytic reactions,\cite{Yang:nat08} although it remains controversial whether the reconstructed or the unreconstructed anatase (001) surface is the most active phase.\cite{Sencer:jpcc13} Well-ordered, ultrathin silica structures consisting of SiO$_4$ units have also been reported \cite{Shaikhutdinov:am13} these bear resemblance to the two-dimensional network on the SrTiO$_3$(110)-($n$~$\times$~1)($n$ = 3 - 6) and rutile TiO$_2$(100)-$c$(2~$\times$~2) surfaces.\cite{Enterkin:natm10, Li:prl11, Warschkow:prl08} It should be noted, however, that Ti in bulk TiO$_2$ and SrTiO$_3$ is octrahedrally coordinated, in contrast to SiO$_2$, which forms tetrahedra also in the bulk.

In this article we present a combined experimental [Scanning Tunneling Microscopy (STM), photoemission spectroscopy (PES)] and theoretical [density functional theory (DFT)] investigation of water adsorption on stoichiometric and reduced SrTiO$_3$(110) surfaces with a two-dimensional tetrahedrally-coordinated (4~$\times$~1) reconstructed layer. Both experimental and theoretical results clearly show that water dissociates on the surface with oxygen vacancies (V$_{\textrm O}$'s), while water neither adsorbs nor dissociates on the stoichiometric surface at room temperature (RT).  Generalizing our result we propose that two-dimensional, tetrahedrally coordinated overlayers on oxide materials interact only weakly with water.

\section{2. Materials and Methods}

\subsection{2.1 Experimental Details}

STM measurements were performed in two ultra-high vacuum (UHV) chambers equipped with a SPECS Aarhus STM at RT and an Omicron low temperature (LT) STM at 78 K, respectively (see Refs \citen{Parkinson:jacs11} and \citen{Scheiber:prl10} for more details).  Synchrotron radiation photoemission spectroscopy experiments were performed at beamline I311 at the MAX IV Laboratory.\cite{Nyholm:nucl01} The pressure in all UHV systems was better than 1~$\times$~10$^{-10}$ mbar.  Nb-doped (0.5 wt\%) SrTiO$_3$(110) single crystals were purchased from MaTeck, Germany.  The clean surface was prepared by cycles of Ar$^+$ sputtering (1 keV, 5 $\mu$A, 10 minutes) and followed by annealing in 2~$\times$~10$^{-6}$ mbar oxygen at 900 $^\circ$C for 1 h.\cite{Wang:apl09} The samples were heated by electron bombardment (13 mA, 900 V) or by passing alternating current through the crystal, and the temperature was monitored with an infrared pyrometer.  The surface reconstruction was checked by low energy electron diffraction (LEED) and was adjusted by depositing Sr or Ti on the surface at RT followed by annealing until a sharp (4~$\times$~1) LEED pattern was observed.\cite{Wang:prb11} The surface was exposed to atomic H by backfilling the chamber with H$_2$ while keeping a hot tungsten filament in line of sight with the sample. The hydrogen cracking efficiency in our setup is estimated 5\% with the W filament temperature about 2000 $^{\circ}$C.\cite{Atsushi:jjap95} The density of H atoms is around 0.1 per nm$^2$ after dosing at a H$_2$ partial pressure of 1$\times$10$^{-6}$ mbar for 5 min with the sample at room temperature. Deionized H$_2$O was cleaned by repeated freeze-pump-thaw cycles and dosed by backfilling the UHV chamber through a leak valve.  The purity of the water vapor was checked by mass spectrometry. All photoemission spectra in this paper were collected with the emission normal to the sample plane; the angle between the sample normal and the incoming x-rays was 54.7$^\circ$. Photon energies were 605 eV and 45 eV for core-level and valence band photoemission spectroscopy, respectively. The binding energies were calibrated with respect to the Fermi level of a clean Mo sample plate, on which our sample was mounted.

\subsection{2.2 Computational Details}

The first-principles calculations were performed using the projector augmented-wave method as implemented in the Vienna \textit{ab initio} simulation package (VASP) code,\cite{vasp1, vasp2} using the Perdew-Burke-Ernzerhof (PBE)\cite{pbe} approximation to treat the exchange-correlation functional within the DFT.  The kinetic energy cutoff for the plane waves expansion was set to 600 eV, and reduced to 400 eV for the nudged elastic band (NEB) calculations as detailed below.  In order to improve the description of dispersion forces, which are expected to play an important role in H$_2$O physisorption phenomena and are not correctly accounted for in standard DFT, we have employed two alternative corrections: (i) the DFT-D2 method of Grimme\cite{Grimme:jcc04, Grimme:jcc06, Grimme:obc07} and (ii) the modified version of van der Waals DFT (vdW-DFT), adopting the recently introduced functional optB86b-vdW.\cite{Klimes:prb11}

Our surface calculations are based on the SrTiO$_3$(110)-(4~$\times$~1) structural model proposed by Enterkin \textit{et al.}.\cite{Enterkin:natm10} To weaken the interaction between the water and its periodic image we have adopted a large (4~$\times$~2) supercell (Fig. 1), which is constructed by doubling the (4~$\times$~1) model along the [1$\overline{1}$0] direction. We have used a symmetric slab consisting of 13 layers separated by a vacuum layer of 12 \AA~(total thickness 32 \AA). A pair of H$_2$O molecules was symmetrically adsorbed on both sides of the slab.  During structural optimization all atoms were allowed to relax until all components of their residual forces were less than 0.02 eV/ \AA, except for the atoms in the central three layers, which were kept fixed in their bulk positions. We have used the fully optimized PBE lattice constant, 3.945 \AA~(very close to the corresponding experimental one, 3.905 \AA), and a (2~$\times$~3~$\times$~1) Monkhorst-Pack k-point mesh (reduced to 1~$\times$~1~$\times$~1 for the NEB runs) for the Brillouin-zone integrations.  

The oxygen vacancy formation energy E$_{\textrm f}$(V$_{\textrm O}$) is computed as E$_{\textrm f}$(V$_{\textrm O}$)=1/2[E$_{\textrm {TOT}}$(2V$_{\textrm O}$)-E$_{\textrm {TOT}}$+E(O$_2$)] where E$_{\textrm {TOT}}$ refers to the DFT total energy of the clean symmetric slab, E$_{\textrm {TOT}}$(2V$_{\textrm O}$) denotes the DFT total energy of the symmetric slab containing two V$_{\textrm O}$'s, and E(O$_2$) indicates the DFT energy of the oxygen molecule. Similarly, the H and H$_2$O adsorption energies are evaluated using the formula E$_{\textrm {ads}}$(X)=1/2[E$_{\textrm {TOT}}$(2X)-E$_{\textrm {TOT}}$-2E(X)] (with X=H and H$_2$O), where EE$_{\textrm {TOT}}$(X) refers to the DFT total energies of the symmetric slab containing two H adatoms or two water molecules, whereas E(X) represents the DFT energies of the isolated H atom or H$_2$O molecule.

The energy barriers for the water dissociation processes were determined via the climbing image NEB (CI-NEB) method,\cite{Henkelman:jcp00} which is designed to compel one of the intermediate states near the transition point to climb up along the reaction coordinate to reach the highest saddle point, thus leading to a more accurate evaluation of the energy barrier than the regular NEB does. Due to the computational load, we adopted 4-8 images connecting two subsequent minima of the potential energy surface for determining the minimum energy path. The whole path was considered to be converged when the residual forces acting on the individual images dropped below the threshold of 0.05 eV/ \AA. For the NEB calculations we did not include dispersion corrections on top of DFT, as it has been demonstrated that these have little impact on the activation energy.\cite{Sorescu:jcp11}

\section{3. Results}

\subsection{3.1 Scanning Tunneling Microscopy Studies}


Figure 2a shows an empty-states STM image of the SrTiO$_3$(110) surface after exposure to atomic hydrogen at an H$_2$ pressure of 1$\times$10$^{-6}$ mbar for 5 min.  The bright stripes along the [1$\overline{1}$0] direction correspond to the Ti(III) and Ti(II) atoms in the six-membered rings, located in tetrahedral units that connect to the SrTiO$_3$ substrate below by sharing corners. The ridges are separated by a dark trench originating from the tetrahedra in the ten-membered rings, which share edges with the SrTiO$_3$ underneath (see Fig. 1).  Each stripe contains two or three bright rows of periodic dots for the (4~$\times$~1) or (5~$\times$~1) reconstruction, respectively.\cite{Li:prl11} On top of the stripes, two types of bright protrusions are observed.  Sr adatoms, which are part of the (4~$\times$~1) antiphase domain structure\cite{Wang:prl13} are labeled with red arrows. In agreement with the DFT calculations\cite{Wang:prl13} they are adsorbed in the middle of the six-membered rings, \textit{i.e.}, centered on  the bright (4~$\times$~1) stripes.  The Sr adatoms have an apparent height of $\sim$240~pm. [Quoted here and in the following are typical values for the apparent heights observed for an STM sample bias of +2.3 V and a tunneling current of 0.1 nA. Note however, the apparent height also dependents on the tip state.] 

It is well accepted that atomic hydrogen preferentially adsorbs on the surface oxygen atoms, forming hydroxyl groups.\cite{Thiel:ssr87, Henderson:ssr02, Weiss:pss02, Diebold:ssr03} In our case, the hydroxyl groups (labeled with white arrows in Fig. 2a) have an apparent height of $\sim$130 pm, less than the Sr adatoms. The OH groups appear preferentially at the sides of both, the (4~$\times$~1) and (5~$\times$~1) stripes. DFT calculations (below) show that atomic hydrogen prefers to adsorb at the O3 site (Fig. 1), and the resulting simulated STM image is consistent with experimental results.\cite{Li:prl11} It should be noted that we also observed indications of H interaction with Sr adatoms; note, \textit{e.g.}, the streaky appearance of the extra-bright Sr atom in Fig. 2a that indicates the presence of an adsorbate. 

After flashing the hydroxylated surface to about 300 $^\circ$C, less bright protrusions with an apparent height of $\sim$70 pm appear (blue arrows in Fig. 2b). From TPD and STM experiments it is often observed that molecular water desorbs from hydroxylated oxide surfaces upon flash-annealing.\cite{Parkinson:jacs11, Thiel:ssr87,Henderson:ssr02} Indeed, from a prior TPD study a similar conclusion was drawn for the SrTiO$_3$(001) surface.\cite{Wang:2002jvsta} It was observed that molecular water desorbs above 100 $^\circ$C on the hydroxylated SrTiO$_3$(001) surface.\cite{Wang:2002jvsta} Therefore, it is reasonable to attribute the less bright protrusions to V$_{\textrm O}$'s. The V$_{\textrm O}$'s sit also at the side of the (4~$\times$~1) stripes, similar to the hydroxyls. These results agree very well with the preference for a V$_{\textrm O}$ at the O3 site in DFT calculations as shown in the following and in Ref. \citen{Li:prl11}.

Figure 3a shows an LT-STM image of the SrTiO$_3$(110) surface after exposure to 0.3 Langmuir (L) water at 110 K.  Bright features with an apparent height of $\sim$80 pm, labeled with green arrows, appear in the trenches between stripes. These features are different from the V$_{\textrm O}$'s and hydroxyls in Fig. 2. From TPD measurements on the SrTiO$_3$(001) surface, molecular water starts to desorb around 200 - 260 K at low exposure (< 1 L), while weakly bound and multilayer water desorbs below 200 K upon further exposure.\cite{Wang:2002jvsta} We attribute the features in Fig. 3a to molecular water that is located at the cation sites at low exposure. From the DFT calculations shown below, molecular water preferentially adsorbs at the TiI site in the ten-membered rings on the SrTiO$_3$(110)-(4~$\times$~1) surface (Fig. 1b), consistent with the experimental observations.(5~$\times$~1) stripes (Fig. 3b), indicating hydroxyl formation after dosing water at RT. In addition to single hydroxyls, hydroxyl pairs are also observed on the surface, again labeled by white arrows in Fig. 3b. These pairs are likely due to the dissociation of water at the V$_{\textrm O}$'s. Note that the distance between these hydroxyl pairs is two unit cells along   direction, indicating a repulsive interaction between them. Here the saturation coverage of hydroxyls is approximately 0.01 ML (1 ML = 4.64~$\times$~10$^{14}$ atoms/cm$^2$ relative to the SrTiO$_3$(110)-(1~$\times$~1) unit cell), suggesting a surface V$_{\textrm O}$ density of half that value. Further increasing the water dosage up to 50 L does not introduce more hydroxyls on the surface, and no indication of molecular water is observed. We conclude that water dissociates only on the V$_{\textrm O}$'s while it neither adsorbs nor dissociates on the stoichiometric surface at RT.

\subsection{3.2 Photoemission Spectroscopy Studies} 

Figure 4 shows photoemission spectra of the valence band region of differently treated SrTiO$_3$(110) surfaces. The valence band of the clean surface shows mainly O 2$p$-derived features. By linearly extrapolating the onset of the spectra, the valence band maximum (VBM) is determined to be located at 3.2 eV below the Fermi level (E$_{\textrm F}$), in agreement with the Nb-doped $n$-type sample and a reported band gap of 3.2 eV for SrTiO$_3$.\cite{Cardona:pr65} For the clean surface no states are observed in the band gap region,\cite{Cao:jcp12} indicating that Nb dopants do not induce in-gap states. This is consistent with the picture that the band structure of lightly $n$-type doped samples can be well described by a simple rigid band shift.\cite{Aiura:ss02}

After dosing up to 240 L water on the clean surface at RT, the valence band spectrum does not change compared to that of the clean surface. For molecularly adsorbed water one would expect features related to its 1b$_2$, 3a$_1$ and 1b$_1$ orbitals.\cite{Valentin:jacs05} On the other hand, an OH 3$\sigma$ state as well as in-gap states can be observed when dissociative adsorption occurs.\cite{Ferrer:1980ss, Webb:1981ss, Egdell:1982cpl, Cox1983247, Eriksen:1987sap, Henrich:prb78} In experiments on as-dosed samples, we did not observe any features related to molecular and dissociative water, in agreement with the conclusion of a rather unreactive surface drawn on the basis of our STM results. 

After dosing atomic hydrogen, an in-gap state with a binding energy of 1.3 eV is observed, as well as a feature state below the O 2$p$ valence band. Partially this feature can be assigned to the OH 3$\sigma$ state, which is located at 10.8 eV.\cite{D'Angelo:prl12} At first sight, the higher binding energy features could be associated with water 1b$_2$ and 3a$_1$ states. However, water does not adsorb on the clean surface at RT, as shown in our STM measurements. Furthermore, no features were observed related to molecular water from the O 1$s$ core-level spectrum for the H-exposed surface (Fig. 5). Instead, STM indicates that H interacts with the Sr adatoms. We tentatively attribute the higher binding energy features to states related to Sr-OH species.\cite{Brookes:1987ssc, Brookes:1987ps} An in-gap state appears after creating V$_{\textrm O}$'s on the clean surface by exposing the surface to intense synchrotron radiation. After exposure to synchrotron light a similar in-gap state and related two-dimensional electron gas were observed on SrTiO$_3$(001) and other perovskite surfaces,\cite{Aiura:ss02, Meevasana:natm11, King:prl12} as well as TiO$_2$ surfaces.50 We find that the in-gap state can be quenched after exposure to O$_2$ at RT, supporting that it arises from V$_{\textrm O}$'s.\cite{Aiura:ss02}

When exposing the surface with V$_{\textrm O}$'s to 1.2 L water at RT, the in-gap state hardly changes. However, a well-defined OH 3$\sigma$ state with a binding energy of 10.8 eV is observed, which indicates water dissociation and formation of hydroxyls. It is well-known that the presence of hydroxyls results in a similar in-gap state as O vacancies.\cite{Valentin:jacs05} This supports the conclusion that water dissociates on the reduced surface. 

Similar conclusions are drawn from the corresponding O 1$s$ core-level photoemission spectra (Fig. 5).  The O 1$s$ spectrum obtained on the clean surface shows a slightly asymmetric peak shape with the main peak located at 530.2 eV and a small shoulder at a higher binding energy of 531.7 eV.  The spectrum does not change after dosing water on the clean surface at RT. After dosing atomic hydrogen and water on the reduced surface, the ratio increases slightly. This result is consistent with observations on titania surface.\cite{Wang:ss95, Ketteler:jpcc07, Walle:prb09}

\subsection{3.3 Electronic Structure Calculations}

To complement the photoemission spectra and achieve an understanding of the electronic properties of the defective and hydroxylated surface as compared to the clean (4~$\times$~1) one, we have determined the most stable configurations and computed their density of states (DOS). By comparing the energies of all possible inequivalent configurations, we determined the most favorable site for the formation of a V$_{\textrm O}$ and for hydrogen adsorption.

The results, collected in Table 1, show that O3 has the lowest V$_{\textrm O}$ formation energy, in agreement with a recent first-principles study\cite{Li:prl11} and consistent with our STM measurements (Fig. 2b). It also represents the most favorable hydrogen adsorption site, with an adsorption energy of 2.19 eV. The most stable hydroxyl is characterized by an O-H bond length of 0.983 \AA, slightly larger than that of a free OH group (0.97 \AA), and a 45.1$^{\circ}$  angle with respect to the surface normal.

We have calculated the DOS of the most favorable oxygen-defective and hydroxylated surfaces. The results are compared to the clean (4~$\times$~1) surface in Fig. 6. Given the well-known drawbacks of standard (local and semilocal) DFT functionals in predicting the correct electronic ground state of strongly correlated electron systems and in describing electron localization effects, we have computed the DOS by means of the PBE+U method,\cite{Anisimov:prb91} using an effective on-site Coulomb repulsion U$_{\textrm {eff}}$ = 4.6 eV for the Ti $d$ states, a choice in line  with previous studies.\cite{Cuong:prl07} The most relevant feature of the V$_{\textrm O}$ case is the appearance of a midgap state right above the valence band maximum, in agreement with the photoemission data.  This state originates from the Ti$^{3+}$ atoms adjacent to the V$_{\textrm O}$, which locally trap the extra electrons created by the V$_{\textrm O}$.  The adsorption of one OH group leads to the formation of only one Ti$^{3+}$, and to the emergence of a feature at about -7 eV below the VBM.  This feature is attributed to the OH-3$\sigma$ bonding state, as shown in the inset of Fig. 6. This picture is reminiscent of the one found for the rutile TiO$_2$(110) surface.\cite{Kurtz:ss89, DiValentin:prl06, Kowalski:prl10}

\subsection{3.4 DFT Calculations: Interaction with H$_2$O}

To elucidate the adsorption of water on the SrTiO$_3$(110)-(4~$\times$~1) surface and to examine the role of V$_{\textrm O}$'s we have performed NEB calculations. One important question is whether the water is predicted to adsorb molecularly or dissociatively on the SrTiO$_3$(110) surface. To answer this question we have investigated the energetics of different adsorption configurations at low coverage, both in molecular and dissociated form, as well as the dissociation energy barriers/pathways among the different configurations. We first focus on the interaction between water and the clean (4~$\times$~1) surface and then we discuss the results obtained for the reduced surface.

\subsubsection{3.4.1 Ideal Surface + H$_2$O}

\textit{\textbf{Molecular adsorption.}} Our first concern is to identify locally stable molecular H$_2$O configurations. We have scrutinized several possible adsorption sites at a coverage of 1/8 ML [one H$_2$O molecule per (4~$\times$~2) unit cell].

The most favorable adsorption site is located in the ten-membered rings near TiI, as shown in the insets of Fig. 7.  The distance between TiI and the water oxygen atom (O$_{\textrm W}$) is found to be 2.341, 2.325 and 2.311 \AA~with the PBE, DFT-D2 and vdW-DFT functional, respectively.  The corresponding water adsorption energies E$_{\textrm {ads}}$(H$_2$O) are -0.716, -1.014 and -1.073 eV, respectively.  As expected, the van der Waals correction substantially increases the magnitude of the adsorption energy, although the geometries are similar to the standard PBE case.  Moreover, the other configurations considered are less stable by 0.15 - 0.5 eV.  Both the H-O$_{\textrm W}$ bond length (1.00 \AA) and the H-O$_{\textrm W}$-H bond angle (106$^{\circ}$) are almost identical to the corresponding values in the free water molecule, 0.985 \AA ~and 104.96$^{\circ}$, respectively. We also evaluated adsorption energies at the experimental condition (300K and 10$^{-9}$ atm) within the framework of \textit{ab initio} thermodynamics.\cite{Reuter:2001prb, Stull:71} The corresponding values are +0.482, 0.184 and 0.125 eV with the PBE, DFT-D2 and vdW-DFT, respectively. The positive value indicates that water does not adsorb on the ideal surface, in agreement with experiment.  

\textit{\textbf{Dissociative adsorption.}} To explore the dissociative configuration (coadsorption of H and OH species), which serves as a basis for studying the water dissociation process, we assumed that the OH species preferentially adsorbs on the Ti atom, and the H atom on the neighboring/next-neighboring surface O atoms. This assumption is reasonable, as no local minimum corresponding to an adsorption at the surface Ti site was found for the H atom.

Most of the dissociative adsorption configurations we explored are unstable (\textit{i.e.} with positive adsorption energy), or relax to the molecular pattern. We established only five stable/metastable dissociative patterns with negative/zero adsorption energy. As mentioned before, here we performed the calculations with the PBE functional, since application of DFT-D2 and vdW-DFT does not alter the adsorption sequence and the geometries. The computed adsorption energy for the most stable pattern is -0.779 eV, about 60 meV more stable then the molecular adsorption case. In the latter configuration (not shown), the OH species anchors on the bridge site between the two Ti surface atoms (TiII and TiIII), while the atom O3, bonded to another H atom, shifts downward due to the electrostatic potential repulsion; this results in two five-fold coordinated Ti atoms. 

On the basis of the computed adsorption energies alone we cannot unambiguously determine whether water molecules are predicted to adsorb molecularly or dissociatively on the SrTiO$_3$(110)-(4~$\times$~1) surface. We have conducted a series of CI-NEB calculations in order to model the dynamics. 

\textit{\textbf{Dissociative reaction.}} We have determined the energy barrier for the water dissociation processes from the most stable molecular adsorption state (initial state) to the geminate dissociative state (final state) by using the CI-NEB method. This procedure allows us to find the minimum energy reaction paths. This pathway choice is appropriate when the strongest molecular adsorption is considered, as in the present case. The resulting energy profile together with the representative structural models is shown in Fig. 7.  The transition barrier for the H$_2$O to dissociate on the SrTiO$_3$(110)-(4~$\times$~1) surface is rather large (>~1.6~eV), much higher than the adsorption energies of both the molecularly or dissociated state. This clearly shows that the H$_2$O molecule is not predicted to dissociate on the defect-free surface, in agreement with the experimental observations.

\subsubsection{3.4.2 Reduced Surface + H$_2$O}

As mentioned in the experimental section, significant amounts of hydroxyls are found on the SrTiO$_3$(110)-(4~$\times$~1) surface with V$_{\textrm O}$'s after dosing water. This suggests that the oxygen deficient surface is active with respect to water dissociation. 

The water adsorption energies computed within PBE, DFT-D2 and vdW-DFT are listed in Table 2. All three methods yield very similar values of 1.7 eV, substantially larger (by about 1 eV) than those on the stoichiometric, non-defective surface. The adsorption of water on the defective surface is clearly favorable. Van der Waals interactions do not play a significant role, which is suggestive of a primarily chemisorption process. 

Considering that the three different methods also deliver a quantitatively similar description of the structural characteristics (see Table 2), we will focus on the PBE results only in the following. The structural model of the optimized initial configuration is provided in Fig. 8.  In the optimized structure, the H$_2$O molecule is slightly tilted towards one of the threefold-coordinated Ti atoms near the V$_{\textrm O}$, forming two asymmetric Ti-O$_{\textrm W}$ bonds of 2.107 and 3.090 \AA. One of the H-O$_{\textrm W}$ bonds in the adsorbed water molecule points towards the O4 atom forming an H-bond with a bond length of 1.571 \AA, in turn slightly enlarging the molecular H-O$_{\textrm W}$ bond length to 1.055 \AA. The second H remains free, connected to the O$_{\textrm W}$ with the corresponding H-O$_{\textrm W}$ bond length (0.985 \AA).

As aforementioned, we primarily focus on the original geminate dissociative states.  The OH species occupies the O3 vacancy site, with the remaining H atom anchored to the neighboring O4 atom (see inset in Fig. 8).  The resulting O4-H is nearly flat-lying and is H-bonded with the adjacent O4' surface oxygen atom. This structural and chemical environment results in a large adsorption energy of 2.28 eV. This is already a strong indication that the water molecule is preferentially adsorbed dissociatively rather than molecularly.  However, exothermicity is a necessary, but not sufficient, condition for dissociation.  In order to gain more insights into the dissociative adsorption process we have conducted NEB calculations for the energy barrier.  The resulting energy profile for the dissociation pathway in Fig. 8 shows an energy barrier at the transition state of 0.9 eV.  This barrier is significantly lower than the corresponding values (1.7 eV) obtained on the ideal surface, clearly indicating that V$_{\textrm O}$'s strongly facilitate water dissociation.  This is again in excellent agreement with the experimental observations, which reveal that water interacts with V$_{\textrm O}$'s, forming two hydroxyl groups on the surface.  Similar energy pathways for the dissociative process were also found for the defective surface with an O4 vacancy, which is characterized by an exothermic energy of 1.4 eV and a slightly larger barrier of 1.1 eV. Given the theoretical and experimental results above, it is clear that V$_{\textrm O}$'s facilitate water dissociation on the SrTiO$_3$(110) surface and the barriers are low enough for this process to happen at room temperature or slightly above RT.

\section{4. Discussion}

Our DFT calculations show that the V$_{\textrm O}$'s are preferentially created at the O3 site in the six-membered ring of the (4~$\times$~1) reconstruction, which is also the most faV$_{\textrm O}$rable site to form OH. This is in excellent agreement with the experimental STM images (Fig. 2). Moreover, the water tends to adsorb molecularly at the TiI site in the ten-membered ring, where it appears as bright protrusions between the stripes in the STM images at low temperature (Fig. 3a).

The situation is different when water is dosed in the presence of V$_{\textrm O}$'s.  Both the experimental and theoretical results unequivocally show that water dissociates spontaneously at the V$_{\textrm O}$'s at finite temperatures.  In STM the two OH groups resulting from a dissociated water molecule were observed far away from each other. The DFT calculated energy barrier for the direct H diffusion is $\sim$1.35 eV, indicating that the direct hopping is not possible at RT.  Possibly these OH groups are driven apart via the water-assisted mechanism reported in Refs [\citen{Merte:sc12, Wendt:prl06}].

Overall, the ideal, non-defective SrTiO$_3$(110)-(4~$\times$~1) surface is remarkably inert towards water adsorption, while the V$_{\textrm O}$'s facilitate bonding and dissociation of water.  Moreover, V$_{\textrm O}$'s created on the SrTiO$_3$(110) surface are metastable, inclined to diffuse to subsurface sites as suggested in previous studies.\cite{Li:prl11} Therefore, under real-world conditions we expect vacancy-mediated adsorption and dissociation to be rare on this surface. 

As stated in the introduction section, SrTiO$_3$(110) is a polar surface, consisting of alternating (SrTiO)$^{4+}$ and (O$_2$)$^{4-}$ planes in the bulk. While an uncompensated polar surface is unstable and chemically active, our results indicate that, in this case, compensating polarity with the reconstruction network is very efficient in creating an inert surface.  The most peculiar structural feature of the reconstruction is the presence of the TiO$_4$ tetrahedra on the top layer.  Interestingly, the TiI-tetrahedra (edge-sharing with the substrate) in the ten-membered rings are reminiscent of a similar configuration at the reconstructed anatase TiO$_2$(001)-(1~$\times$~4) surface,\cite{Lazzeri:prl01} which contains a distorted TiO$_4$ tetrahedron.  It was demonstrated that water dissociates spontaneously on the ridge of this reconstructed surface in theoretical and experimental studies.\cite{Gong:jpcb06, Blomquist:jpcc08} In fact, the anatase TiO$_2$(001) surface is considered the most active facet in photocatalytic reactions.\cite{Yang:nat08} An analysis of our results gives insights as to why the TiO$_4$ tetrahedra are so inert in the case of SrTiO$_3$(110)-(4~$\times$~1).

\textit{\textbf{Electronic Aspects.}} While excess electrons located at the energies near the band gap of reducible oxide surfaces are generally connected with a high reactivity,\cite{Lu:jpc94} the clean SrTiO$_3$(110)-(4~$\times$~1 ) surface has no in-gap states in both, experiment and theory. In fact, an analysis of the layer-resolved DOS (not shown) indicates that the top layer has a slightly larger band gap, compared to the SrTiO$_3$ layers underneath.  In spite of the 4-fold coordination in this tetrahedral configuration, the Ti atom should not be considered undersaturated. The Ti atom hybridizes with the four surrounding oxygen atoms, forming strong covalent bonds with a short bond length, which lead to the relatively large band gap.  Experimental and theoretical results also show that the Ti valence is 4+ and no in-gap state is present on the anatase TiO$_2$(001)-(1~$\times$~4) surface.\cite{Lazzeri:prl01, Chambers:ss09} Thus, while explaining our inert SrTiO$_3$(110)-(4~$\times$~1) surface, the electronic structure provides no argument for the supposedly reactive TiO$_4$ tetrahedra on anatase.

Interestingly, similar tetrahedrally coordinated TiO$_4$ units are present on the TiO$_2$(110)-(1~$\times$~2) surface, forming one-dimensional Ti$_2$O$_3$ rows.\cite{Blanco-Rey:prl06} Due to the presence of Ti$^{3+}$ species the TiO$_2$(110)-(1~$\times$~2) surface is proposed to be chemically active, as demonstrated by reacting with NO.\cite{Abad:langmuir07} It would be interesting to test whether this surface is also reactive for water dissociation.

\textit{\textbf{Structural Aspects.}} What is needed for strong water interaction are freely accessible acidic sites, and a neighboring O atom that can act as Br\o nsted base (proton acceptor). In our case, the TiO$_4$ tetrahedron is quite regular: the bond length ranges from 1.826 to 1.896 \AA, with a O-Ti-O bond angle range of 92.82 -123.08$^{\circ}$. The acidic Ti sites are significantly recessed into the surface compared to the surrounding oxygen atoms, making them inaccessible and non-reactive.  In contrast, on the anatase TiO$_2$(001)(1~$\times$~4) reconstructed surface, the TiO$_4$ tetrahedron is very distorted; the bond length along the [100] direction consists of alternating long (2.134 \AA) and short (1.831 \AA) Ti-O bonds, while the bonds along the [010] directions are identical (1.805 \AA).  The O-Ti-O bond angle along the [100] and the [010] direction is 145.15$^{\circ}$ and 104.76$^{\circ}$, respectively,\cite{Gong:jpcb06} leading to the exposure of the Ti atom as an active acidic site. Furthermore, and at variance with what was found for SrTiO$_3$(110)-(4~$\times$~1), the distorted TiO$_4$ tetrahedron on the anatase TiO$_2$(001)(1~$\times$~4) surface forms a quasi-one dimensional row along the [100] direction. This flexible framework provides the freedom of relaxation, and facilitates the water dissociation.  At the SrTiO$_3$(110)-(4~$\times$~1) surface, the two-dimensional nesting of the six- and ten-membered rings is more rigid, which contributes to its inertness.

A similar two-dimensional reconstructed overlayer consisting of corner-sharing TiO$_4$ regular tetrahedra has been established on the rutile TiO$_2$(100)-$c$(2~$\times$~2) surface.\cite{Warschkow:prl08} From the present results, we would expect this reconstructed surface also to be relatively inert; it would be interesting to test this prediction.

\section{5. Summary and Conclusion}

We have performed a systematic study of water interaction with the two-dimensional titania overlayer consisting of TiO$_4$ tetrahedra, on the SrTiO$_3$(110)-(4~$\times$~1) surface with and without oxygen vacancies.  We found that water dissociates on the oxygen vacancies, in line with many other oxide surfaces.  We also found the two-dimensional, tetrahedrally coordinated TiO$_4$ overlayer to be remarkably inert, in contrast to the one-dimensional, tetrahedrally coordinated TiO$_4$ units at the anatase TiO$_2$(001)-(1 $\times$4) surface.  The weak water adsorption on this surface stems from the regular tetrahedra and the two-dimensional rigid network, as well as its insulating electronic structure. Recently, TiO$_4$ tetrahedra have emerged as a common building block on many Ti-containing oxides surfaces.  We expect that our conclusions of a inert two-dimensional top layer should also apply to these newly-discovered surfaces.

\acknowledgement

This work has been supported by the Austrian Science Fund (FWF) under Project No. F45 and the ERC Advanced Research Grant `OxideSurfaces'. E.M. acknowledges support from the FWF under Project No. W1243 (Solids4Fun). All DFT calculations were performed at the Vienna Scientific Cluster (VSC-2). Valuable discussions with Annabella Selloni and Laurence Marks are gratefully acknowledged.



\providecommand*\mcitethebibliography{\thebibliography}
\csname @ifundefined\endcsname{endmcitethebibliography}
  {\let\endmcitethebibliography\endthebibliography}{}

 \begin {figure}[b]
 \includegraphics [width=3.2 in,clip] {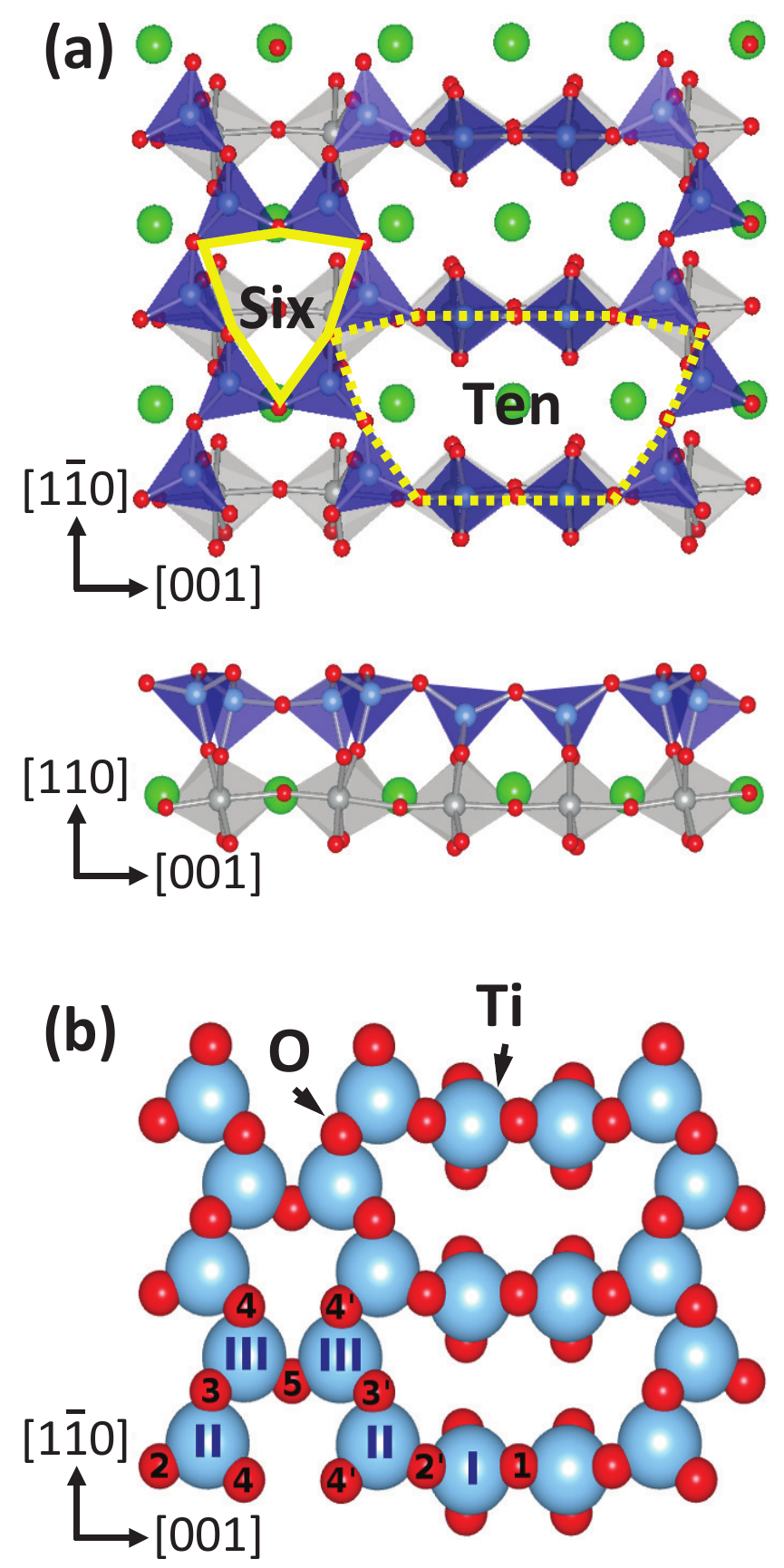}
 \caption{
 Model of the SrTiO$_3$(110)-(4~$\times$~1) surface. (a) Top and side views. The reconstructed layer consists of a network of darker TiO$_4$ tetrahedra (blue) forming six- and ten-membered rings, on top of the SrTiO$_3$(110) substrate, which contains TiO$_6$ octahedra (lighter, gray). Large, medium and small spheres denote Sr, Ti and O atoms, respectively. (b) Top view of the topmost reconstructed layer with the surface Ti and O atoms labels used in the present study. 
}
\label{Fig1}
\end{figure}

 \begin {figure}[b]
 \includegraphics [width=3.2 in,clip] {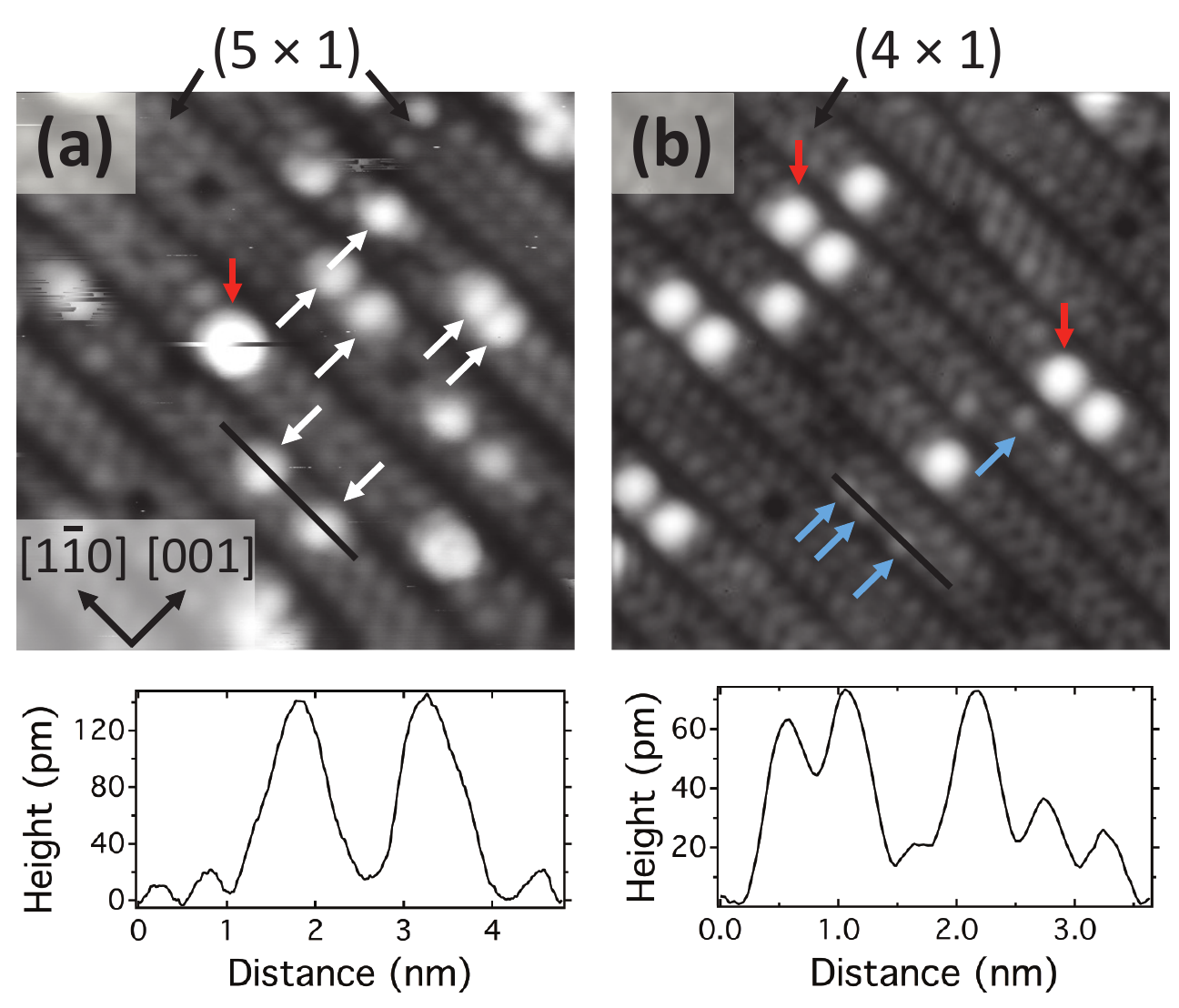}
 \caption{
STM images (image size 9~$\times$~9 nm$^2$, sample bias +2.3 V, tunneling current 0.1 nA) of the SrTiO$_3$(110) surface. The surface exhibits an overall (4~$\times$~1) reconstruction; locally a few (5~$\times$~1) rows are apparent. (a) After exposure to atomic hydrogen and (b) after flashing the surface in (a) to $\sim$300 $^{\circ}$C. Sr adatoms, hydroxyls and oxygen vacancies appear in various levels of brightness and are labeled by red, white and blue arrows, respectively. The line profiles in the lower panels were taken at the lines shown in the STM images.
}
\label{Fig2}
\end{figure}

 \begin {figure}[b]
 \includegraphics [width=3.2 in,clip] {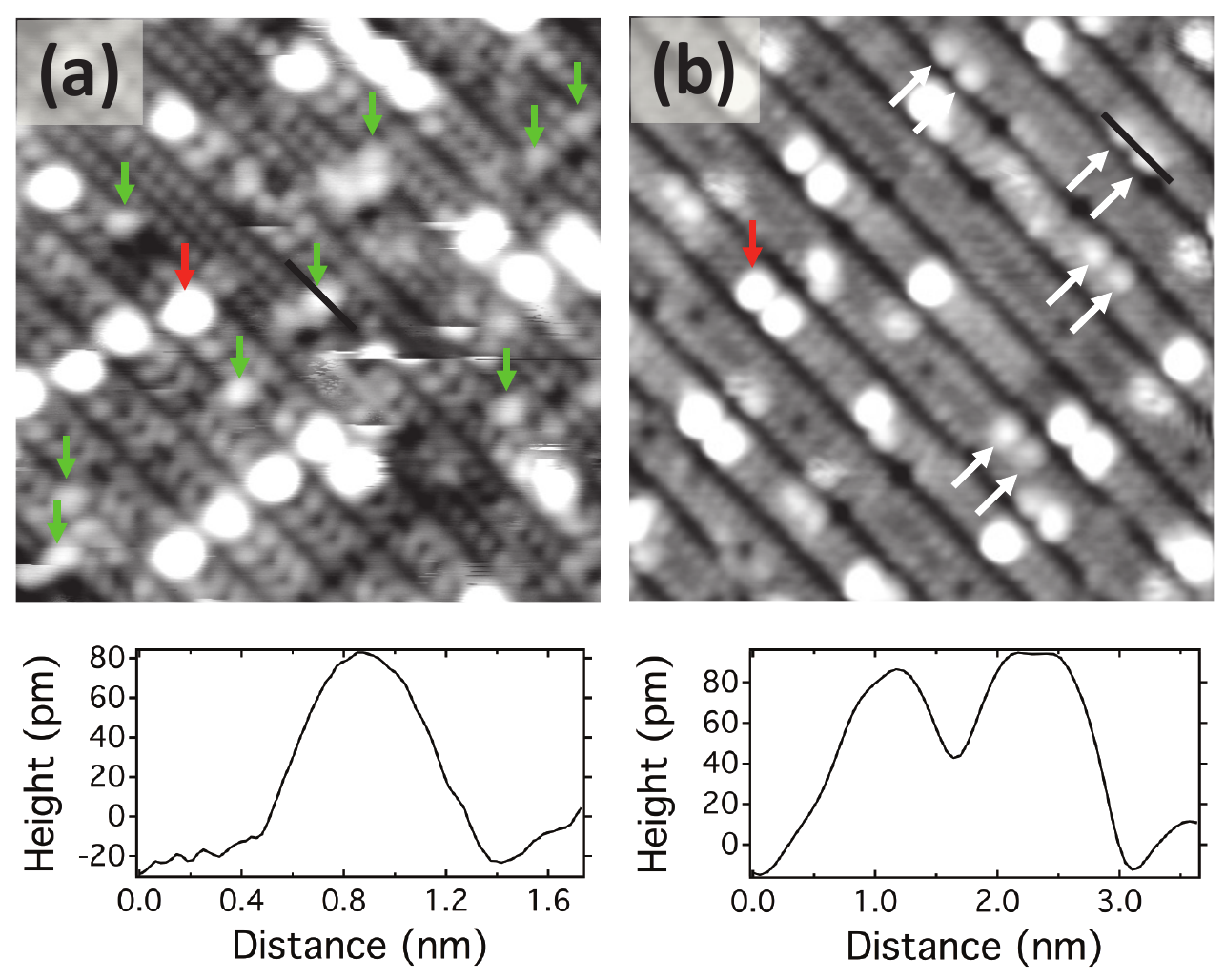}
 \caption{
STM images (18~$\times$~18 nm$^2$, 2.1 V, 0.1 nA) of the SrTiO$_3$(110)-(4~$\times$~1) surface after exposure to (a) 0.3 L water at 110 K, imaged at 78 K; (b) 3L at RT, imaged at RT. Green and white arrows point to molecular water and hydroxyl pairs, respectively. As in Fig. 2 the red arrows point out single Sr adatoms. The line profiles in the lower panels were taken at the lines shown in the STM images.
}
\label{Fig3}
\end{figure}

 \begin {figure}[b]
 \includegraphics [width=3.2 in,clip] {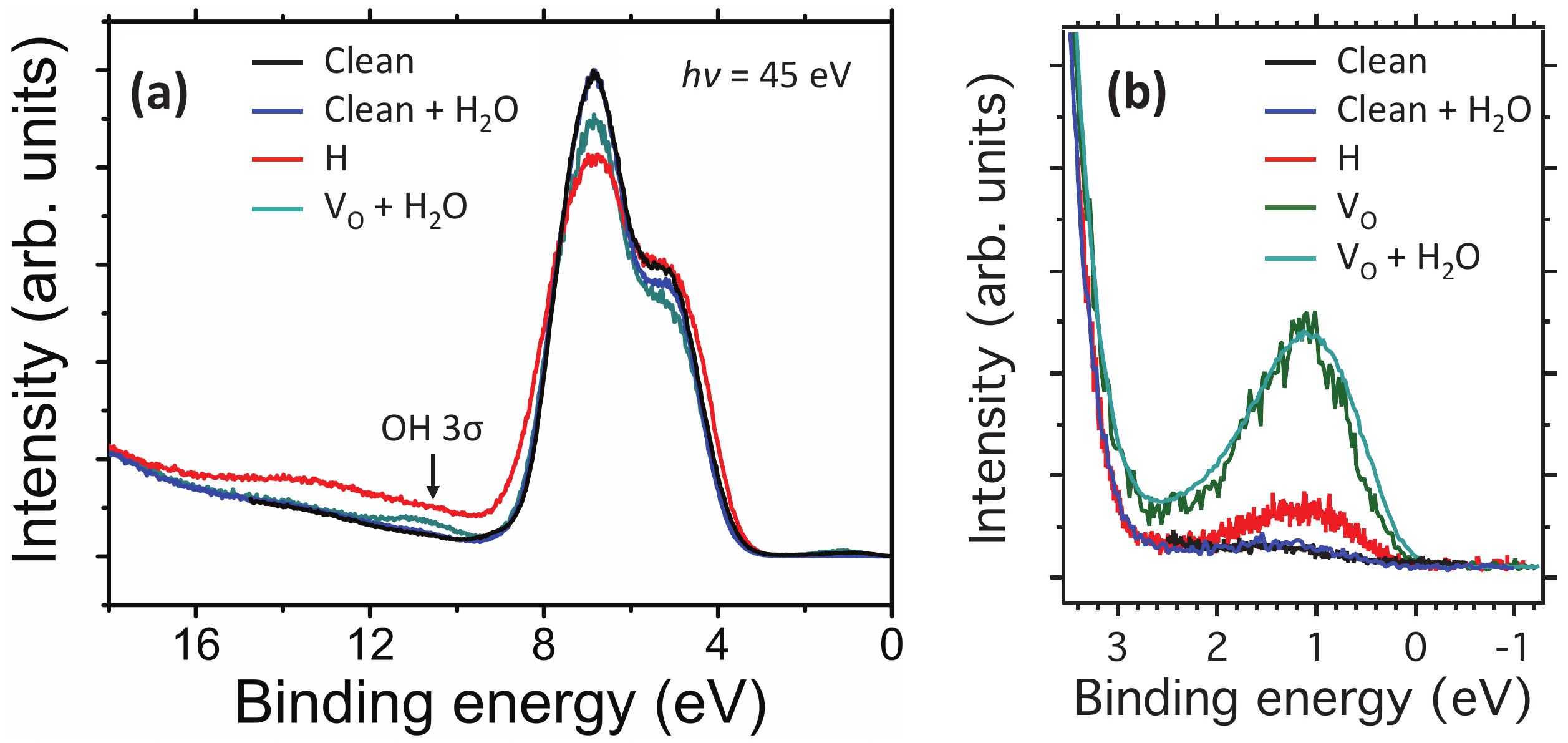}
 \caption{
Comparison of valence band photoemission spectra of the clean surface (black), after exposure to water (blue), atomic hydrogen (red), of a surface with oxygen vacancies (green), and after exposure to water (cyan). All spectra were taken at RT.
}
\label{Fig4}
\end{figure}

 \begin {figure}[b]
 \includegraphics [width=3.2 in,clip] {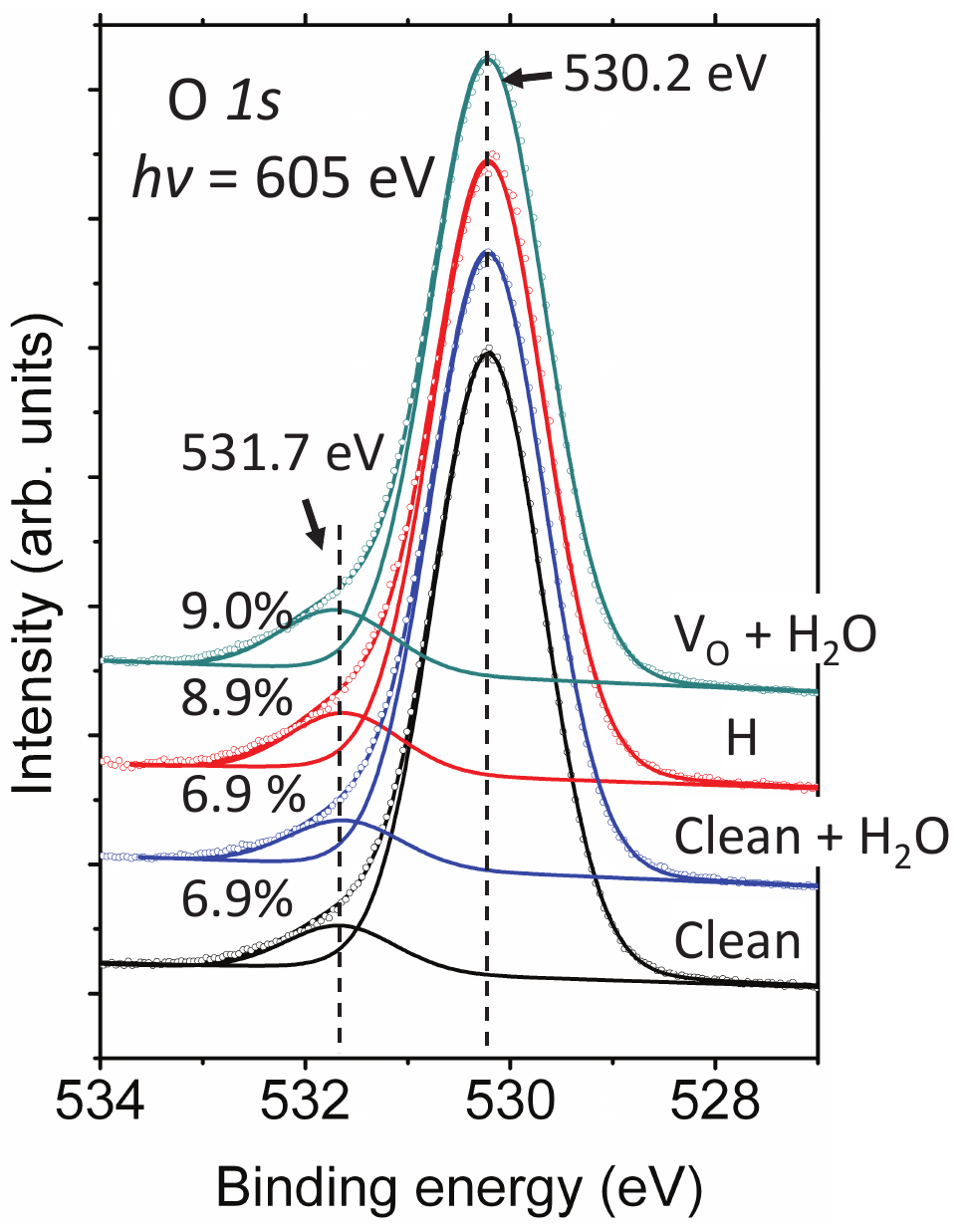}
 \caption{
Comparison of O 1$s$ core-level XPS spectra of the clean surface (black), after exposure to water (blue) and atomic hydrogen (red), and surface with oxygen vacancies exposed to water (cyan). All spectra were taken at RT.
}
\label{Fig5}
\end{figure}

 \begin {figure}[b]
 \includegraphics [width=3.0 in,clip] {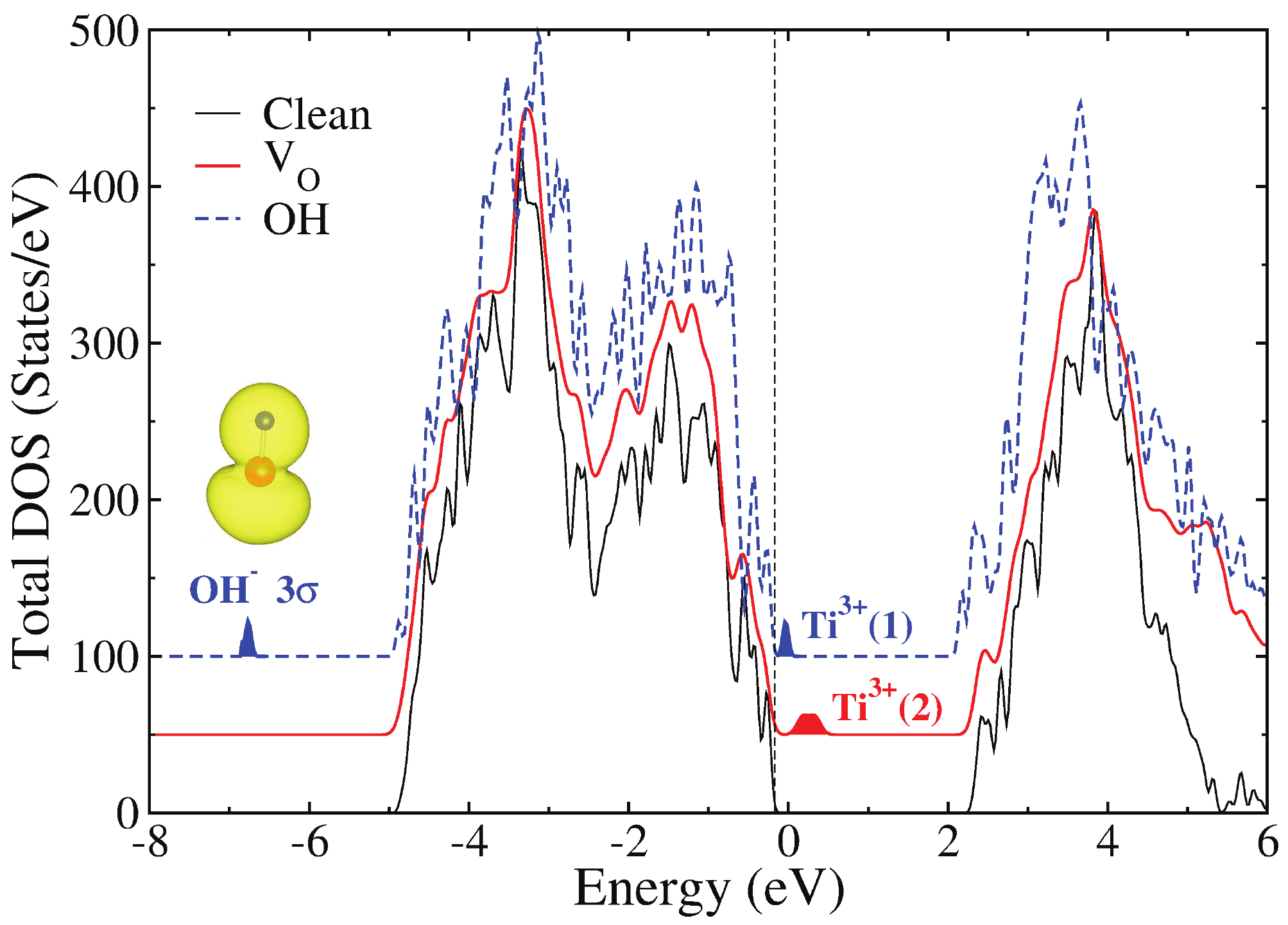}
 \caption{
PBE+$U$ valence and conduction band density of states of the clean SrTiO$_3$(110)-(4~$\times$~1) surface (thin black line) and the reduced surface with an oxygen vacancy (red full line) and hydroxyl species (dashed blue line). All spectra are aligned with respect to their valence-band maxima. The Ti$^{3+}$ midgap states [both singly, Ti$^{3+}$ (1), and doubly, Ti$^{3+}$ (2) occupied] as well as the OH-3$\sigma$ states are highlighted with a colored background.
}
\label{Fig6}
\end{figure}

 \begin {figure}[b]
 \includegraphics [width=3.2 in,clip] {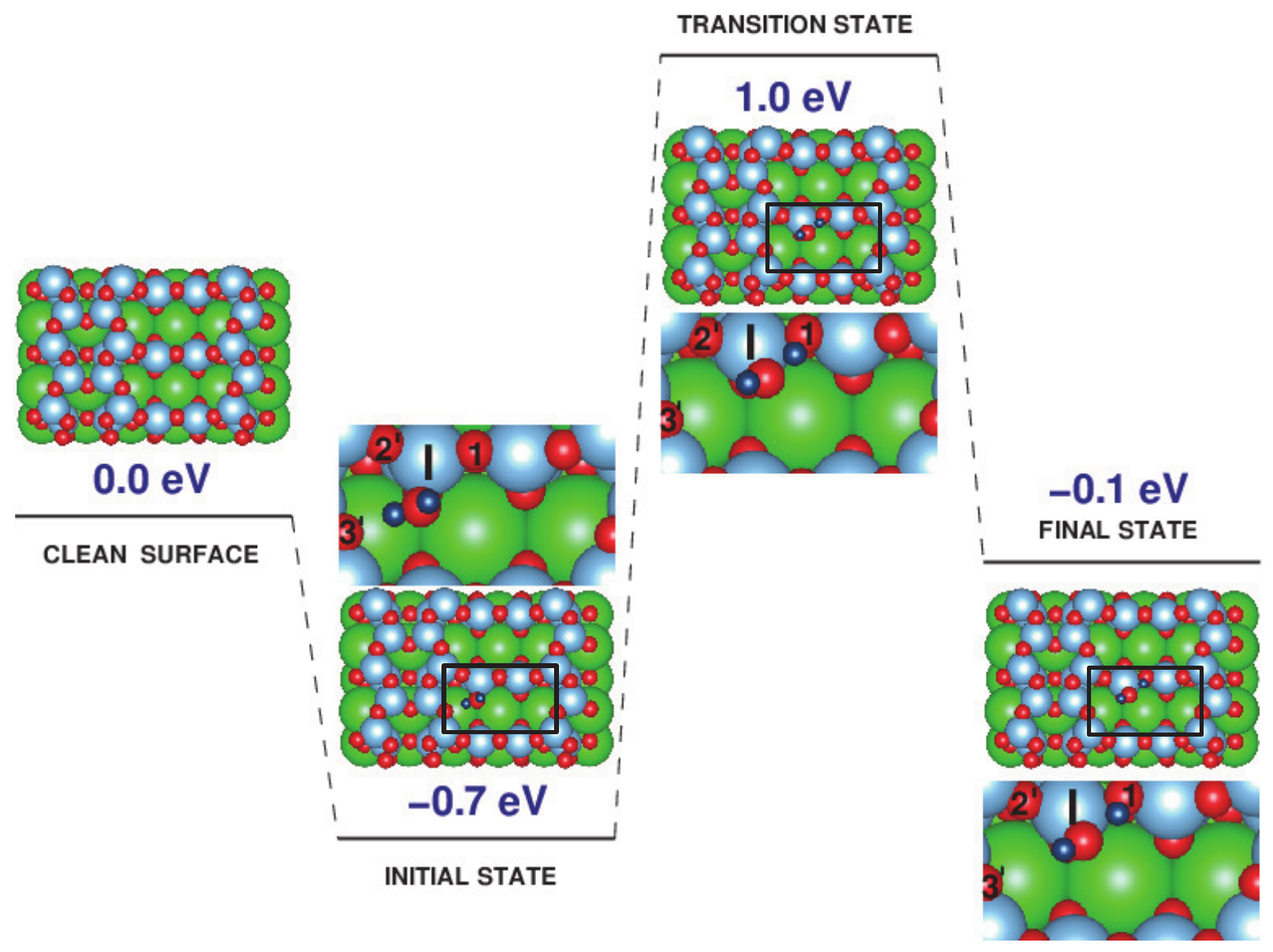}
 \caption{
Potential-energy profile for the reaction of an adsorbed H$_2$O molecule on the ideal, non-defective SrTiO$_3$(110)-(4~$\times$~1) surface. The energy zero corresponds to the H$_2$O in the gas phase far away from the surface. For each state considered the corresponding optimized structures are shown as insets in wide and zoomed view.
}
\label{Fig7}
\end{figure}

 \begin {figure}[b]
 \includegraphics [width=3.2 in,clip] {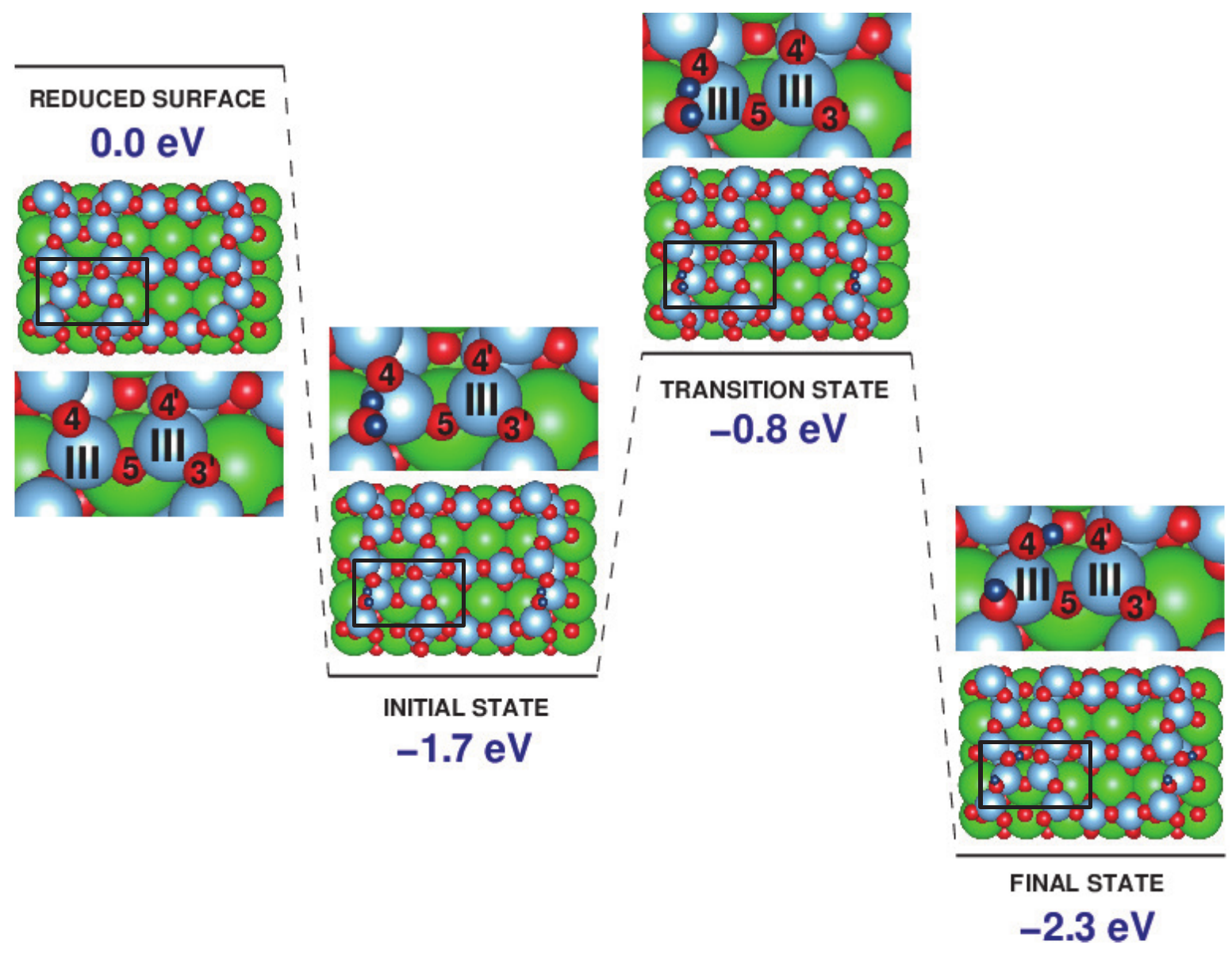}
 \caption{
Potential-energy profile for the reaction of an adsorbed H$_2$O molecule at the defective SrTiO$_3$(110)-(4~$\times$~1) surface. The energy zero corresponds to the H$_2$O in the gas phase, far away from the surface.
}
\label{Fig8}
\end{figure}


\break

\begin{table}[b]
\caption{Oxygen vacancy formation energy E$_{\textrm f}$(V$_{\textrm O}$) and hydrogen adsorption energy E$_{\textrm {ads}}$(H) (in eV) for different oxygen sites (following the labeling given in Fig. 1b) obtained with the PBE functional. Numbers in brackets refer to the relative energy with respect to the most stable configuration. For geometries of adsorbed H see the Supplement.}
\begin{tabular}{c c c c c c}
\hline
\hline
  			  & O1 & O2 & O3 & O4 & O5 \\
\hline
V$_{\textrm O}$ formation energy (eV) & 6.43 & 5.95 & 5.60 & 5.68 & 5.76 \\
			 & (0.83) & (0.35) & (0.0) & (0.08) & (0.16) \\
\hline
H adsorption energy (eV) & 1.79 & 1.93 & 2.19 & 2.16 & 1.62 \\
			 & (-0.40) & (-0.26) & (0.0) & (-0.03) & (-0.57) \\
\hline
\hline
\end{tabular}
\label{table2}
\end{table}

\begin{table}[b]
\caption{Calculated water adsorption energies E$_{\textrm {ads}}$(H$_2$O) (in eV), bond lengths (\AA) (O$_{\textrm W}$ and O$_{\textrm S}$ denotes O atoms in the water molecule and surface, respectively), and H-O$_{\textrm W}$-H angles ($^{\circ}$) for molecular adsorption configurations on the defective surface, calculated with different functionals.}
\begin{tabular}{c c c c c c}
\hline
\hline
Functional & E$_{\textrm {ads}}$(H$_2$O) & Ti-O$_{\textrm W}$ & H-O$_{\textrm W}$ & H-O$_{\textrm S}$ & H-O$_{\textrm W}$-H \\
\hline
\hline
PBE & -1.732 & 2.107, 3.090 & 1.055, 0.985 & 1.571 & 109.22 \\
DFT-D2 & -1.857 & 2.102, 3.061 & 1.057, 0.985 & 1.560 & 109.31 \\
vdW-DFT & -1.727 & 2.098, 3.167& 1.059, 0.987 & 1.552 & 109.35 \\
\hline
\hline
\end{tabular}
\label{table3}
\end{table}

\end{document}